\documentclass{ws-procs10x7}
\usepackage{balance}

\newcolumntype{d}[1]{D{.}{.}{#1}}

\def\Journal#1#2#3#4{{\it #1} {\bf #2}, #3 (#4)}

\makeindex
\begin{document}

\title{BEYOND THE STANDARD MODEL SEARCHES AT THE LHC}

\author{S. WILLOCQ}

\address{Department of Physics, University of Massachusetts, Amherst, MA 01003, USA\\
         E-mail: willocq@physics.umass.edu\\
         Representing the ATLAS and CMS Collaborations}


\twocolumn[\maketitle\abstract{This report presents recent results
 from studies of Beyond the Standard Model physics at the LHC. A focus
 is placed on heavy gauge bosons, electroweak symmetry breaking and left-right symmetry.}
\keywords{Beyond the Standard Model; New Physics; Large Hadron Collider.}
]

\section{Introduction}

  There are many questions left unanswered by the Standard Model (SM)
of particle physics. Many of these revolve around the concept
of symmetries. In this report, the focus is on recent
studies of the prospects for Beyond the Standard Model
searches at the Large Hadron Collider (LHC). Studies in the
context of SUSY, Extra-Dimensions and Black Holes are
reported elsewhere in these proceedings.

\section{Heavy Gauge Bosons}

  The SM is an extremely successful gauge theory
based on the symmetry group $SU(3)_C$ $\otimes$ $SU(2)_L$ $\otimes$ $U(1)_Y$.
Local gauge invariance requires the existence of vector bosons
mediating the strong and electroweak forces: the gluons,
photon, $W$ and $Z$ bosons.
Many extensions of the SM predict the existence of additional
gauge bosons resulting from enlarged symmetry groups.
For example, $E_6$-based theories predict the existence of
additional $U(1)$ groups originating from the breaking of $E_6$
into subgroups, eventually breaking down to the SM group at
low energy.
New heavy gauge bosons are generically refered to as $W^\prime$ and
$Z^\prime$ bosons. These are also predicted in theories with extra
dimensions in the form of Kaluza-Klein excitations of the
electroweak bosons or gravitons.

\begin{figure}[tb]
\vspace{-4mm}
\centerline{\psfig{file=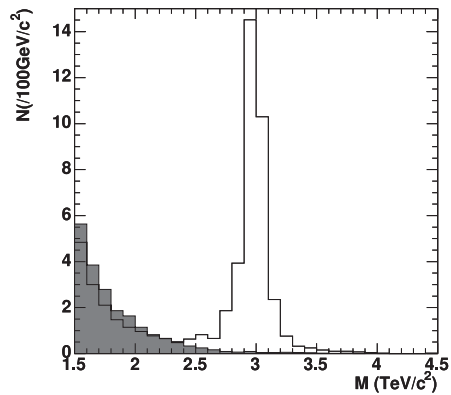,width=2.5in}}
\vspace{-2mm}
\caption{Dielectron invariant mass distribution for a 3-TeV $Z^\prime$ (open histogram)
 and background (shaded histogram), normalized to 30~fb$^{-1}$.}
\label{fig_Zprime_3TeV_eeMass}
\end{figure}

  The ATLAS and CMS Collaborations have studied the potential
for $Z^\prime$ discovery in the $\ell^+\ell^-$ channel (either
$e^+ e^-$ or $\mu^+ \mu^-$)\cite{CMS_TDR,ATLAS_Zprime}.
Dileptons are selected from pairs of isolated electrons or
muons to reduce background from fakes.
Isolation criteria for muons typically rely on tracking only and
do not include energy from the calorimeters since high-momentum muons tend
to lose significant amounts of energy via bremsstrahlung inside the calorimeter.
As an example, the reconstructed di-electron invariant mass distribution
for a 3-TeV Sequential Standard Model (SSM) $Z^\prime$ in CMS is shown
in Fig.~\ref{fig_Zprime_3TeV_eeMass}. In the SSM, $Z^\prime$ bosons couple
to fermions with the same couplings as SM $Z$ bosons.
A very clean peak is observed over a smooth background consisting
primarily of dileptons from the Drell-Yan process
$q\,\overline{q} \rightarrow \gamma/Z^\ast \rightarrow \ell^+\ell^-$.
An integrated luminosity of 0.1--1 fb$^{-1}$ is required for
a $5\,\sigma$ discovery at $M_{Z^\prime} = 1$ TeV, depending
on the assumed model (SSM, $E_6$ or Left-Right symmetric models).
An ultimate reach of $M_{Z^\prime} \sim 5$ TeV is expected
after several years of data taking at design luminosity (100--500 fb$^{-1}$).

  Discrimination between various $Z^\prime$ models can be achieved
by exploiting the fact that these models have different
couplings between the $Z^\prime$ and fermions.
Discriminating observables are:
(i) the total width, which exploits differences in overall coupling strengths,
(ii) the forward-backward asymmetry between positive and negative leptons
relative to the initial quark direction, which exploits differences in parity-violating
couplings (left- vs. right-handed couplings to final state leptons),
(iii) the rapidity of the $Z^\prime$, which exploits differences in the couplings
to the initial state up and down quarks.
The total width observable is most useful for $e^+ e^-$ final states since the
momentum resolution is an order of magnitude worse
for muons than for electrons at these very high momenta.
Most powerful are the total width and the forward-backward asymmetry.

  $W^\prime$ decays have been studied by CMS\cite{CMS_TDR} in the $W^\prime \to \mu \nu$
channel by combining an isolated muon with large missing transverse
energy in the event. The discovery potential is similar to that
for the $Z^\prime$ with an ultimate reach near 6 TeV for 300 fb$^{-1}$.

\section{Electroweak Symmetry Breaking}

  One of the most important goals of the LHC is to elucidate
the origin of Electroweak Symmetry Breaking (EWSB). Two possibilities
need to be considered. Either EWSB is due to a light Higgs boson
or the Higgs boson does not exist and EWSB originates from a new
kind of strong interaction.
In the Higgs boson scenario one needs to cancel quadratic divergences
due to quantum loop corrections to the Higgs mass, to avoid the
well-known fine-tuning problem. The leading candidate for such
a cancelation is Supersymmetry but ``Little Higgs'' models have also
been proposed as an alternative.

\begin{figure}[tb]
\vspace{-4mm}
\centerline{\psfig{file=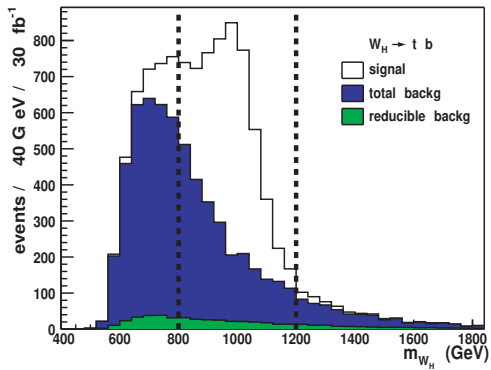,width=2.8in}}
\vspace{-2mm}
\caption{Invariant mass distribution for reconstructed $t\,\overline{b}$
 combinations: $W_H$ signal (open histogram) above background (blue histogram),
 normalized to 30~fb$^{-1}$.}
\label{fig_LittleHiggs_WH_tb}
\end{figure}

  The Littlest Higgs model\cite{lHiggs_model} 
studied by ATLAS\cite{ATLAS_lHiggs,ATLAS_lHiggsHad}
predicts the existence of new heavy gauge bosons ($A_H$, $W_H$ and $Z_H$)
and a heavy $q=+2/3$ quark ($T$) required to cancel quadratic divergences in
the Higgs mass due to quantum loop corrections involving
SM gauge bosons and top quarks.
For masses above 700 GeV, single-$T$ production via
the $W$-exchange reaction $q\, b \to q^\prime\, T$ dominates.
The most promising decay channels are $T \to t\, Z$ (with $t \to b\,\ell\,\nu$)
and $T \to b\, W$ (with $W \to \ell\,\nu$), for which a discovery is expected up to
$M_T =$ 1 and 2 TeV, respectively, with 300 fb$^{-1}$.
Heavy gauge bosons $A_H$, $W_H$ and $Z_H$ can be discovered via their
leptonic decay modes with discovery potential up to about 5 TeV,
depending on the mixing parameter $\cot\theta$ between the
two $SU(2)$ gauge triplets of the model.
However, to discriminate against the many models predicting dilepton resonances,
one needs to also observe decay processes that are typical of
Little Higgs models:
$W_H \to W H$, $Z_H \to Z H$ and $W_H \to t\,\overline{b}$.
The latter process is particularly important if $\cot\theta \sim 1$ since
the first two modes are suppressed in that case.
A study of the process $W_H \to t\,\overline{b}$
with $t \to b \ell \nu$
shows a clear signal excess over background (dominated
by $t\,\overline{t}$ production), see Fig.~\ref{fig_LittleHiggs_WH_tb}.
Discovery of this decay mode is expected up to masses of 3 TeV, depending on
$\cot\theta$.

  Another source of EWSB arises from theories of dynamical EWSB occurring
via new strong interactions. In these theories there is no need for a Higgs
boson, which removes the fine-tuning problem.
CMS studied dynamical EWSB in the context of Technicolor\cite{CMS_TDR}.
In particular, the process $\rho_{TC} \to W Z$, with subsequent decay to
a fully-leptonic final state, has been investigated.
After selection of isolated electrons and muons, measurement of the transverse
missing energy and the application of $W$ and $Z$ kinematical constraints,
an excess of signal events is observed above the background
from $WZ$, $ZZ$, $Z\,b\,\overline{b}$ and $t\,\overline{t}$
production, for the parameters assumed in their study.
%
%

Without a Higgs boson,
the cross section for scattering of longitudinally-polarized
$W$ bosons diverges at high energy, unless new physics contributes to cancel
this divergence.
Such new physics might manifest itself in the form of diboson resonances
at the LHC.
With this motivation, ATLAS studied dynamical EWSB in the general context of the
Chiral Lagrangian model\cite{ATLAS_VBS}. 
In particular, $WZ$ scattering has been investigated in the scattering
process $q_1\, q_2 \to q_1^\prime\, q_2\, W Z$ with the following combination
of leptonic and hadronic final states from the $WZ$ decay:
$\ell\nu \: \ell\ell$, $jj \: \ell\ell$ and $\ell\nu \: jj$.
Events are selected with two forward jets, central leptons or jets
and missing energy (depending on the particular $WZ$ final state).
For $WZ$ final states involving jets,
the event is required not to contain additional central jets and
none of the jets can be identified as $b$-jets.
The main background originates from quark-quark scattering via gluon
and $\gamma/Z$ exchange with the emission of both a $W$ and a $Z$
boson from the outgoing quarks. Other background processes are
$t\,\overline{t}$ and $W$+4 jets.
Intial studies indicate promising sensitivity for diboson
resonances in the $\sim$1 TeV mass region in final states including jets,
for 100 fb$^{-1}$.

A specific experimental issue for this study is the merging of jets
from the decay of high-p$_T$ $W$ or $Z$ bosons, which requires
running jet reconstruction with significantly smaller cones
($\Delta R = 0.2$ instead of 0.7). Further work is also
needed to investigate the impact of pileup on the forward jet tagging.

\section{Left-Right Symmetry}

  Extensions of the Standard Model have been proposed
that explicitly include a left-right symmetry.
Such extensions aim to explain the existence of pure left-handed
charged weak interactions (at low energy) via a spontaneously broken parity.
They also provide a natural explanation for light neutrinos
by incorporating heavy right-handed neutrinos, thereby accommodating
light left-handed neutrinos via the see-saw mechanism.
The left-right symmetric model based on
$SU(2)_L$ $\otimes$ $SU(2)_R$ $\otimes$ $U(1)_{B-L}$
features a triplet ($\Delta^0$, $\Delta^+$, $\Delta^{++}$)
and two doublets of Higgs scalar fields,
and predicts new gauge bosons ($W_R$ and $Z_R$) and
neutrinos ($\nu_R$).

  ATLAS studied\cite{ATLAS_dcHiggs} the striking signal
expected in quark-quark scattering
with the emission of two $W^+$ bosons fusing into
a doubly-charged $\Delta^{++}$ Higgs which then decays
into a pair of like-sign leptons.
The selection requires the presence of two forward jets
and two like-sign electrons, muons or taus.
Main background sources originate from quark-quark scattering
with emission of a $W^+$ boson from each quark, as well as $W t\,\overline{t}$
production.
The discovery for $\Delta^{++}_R$ produced in this mode reaches
up to $\sim$2 TeV for low enough $W_R$ masses, with 300 fb$^{-1}$.

\begin{figure}[tb]
\vspace{-6mm}
\centerline{\psfig{file=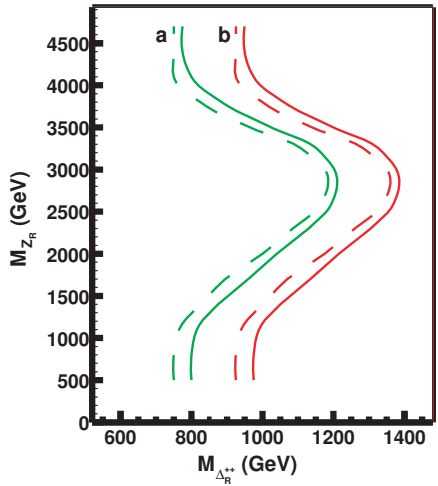,width=2.5in}}
\vspace{-2mm}
\caption{Discovery reach for $\Delta_R^{++}$ Higgs bosons as
 a function of $Z_R$ mass for (a) 100~fb$^{-1}$ and
 (b) 300~fb$^{-1}$. The dashed (solid) line corresponds to the
 requirement of four (three) reconstructed leptons.}
\label{fig_dcHiggs_reach}
\end{figure}

ATLAS also studied the pair production of
$\Delta_{R,L}^{++}\, \Delta_{R,L}^{--}$ in $q\,\overline{q}$
annihilation via virtual $\gamma/Z/Z_{R,L}$.
In this case, a very clean four-lepton
signature is expected and the background is negligible.
The doubly-charged Higgs boson discovery reach,
defined as the observation of 10 signal events,
is shown in Fig.~\ref{fig_dcHiggs_reach}.

\begin{figure}[tb]
\vspace{-3.5mm}
\centerline{\psfig{file=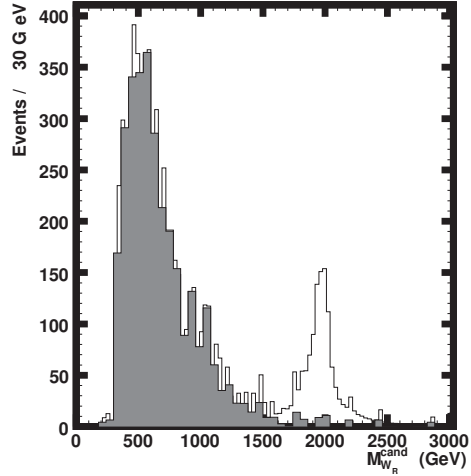,width=2.5in}}
\vspace{-2mm}
\caption{Invariant mass for candidate $W_R \to \ell \ell j j$
 decays with 30~fb$^{-1}$. The shaded histogram corresponds to the
 expected background.}
\label{fig_WR}
\end{figure}

  CMS also studied the left-right symmetric model with a
search for $W_R$ and $\nu_R$\cite{CMS_TDR}. 
These states are produced via $q\,\overline{q} \to W_R \to \ell \nu_R$
and $q\,\overline{q} \to Z_R \to \nu_R \overline{\nu_R}$, with
$\nu_R \to \ell j j$.
Events are selected with at least two isolated electrons and at least
two jets.
A clean $W_R$ signal is observed in the $eejj$ mass distribution
for $M_{W_R} = 2000$ GeV and $M_{\nu_R} = 500$ GeV,
see Fig.~\ref{fig_WR}.

\section{Summary and Outlook}

  ATLAS and CMS are gearing up for data taking at the energy frontier
when the LHC begins operations at $\sqrt{s} = 14$ TeV in 2008.
Both experiments have significant discovery potential in areas
related to fundamental symmetries.
In particular, the discovery reach is
up to $\sim$5-6 TeV for heavy gauge bosons,
$\sim$2 TeV for the heavy top quark in Little Higgs models,
$\sim$600 GeV for the $\rho_{TC}$ technihadron,
$\sim$2 TeV for doubly-charged Higgs bosons
and $\sim$2.5 TeV for heavy neutrinos.

\section*{Acknowledgments}
I'd like to thank the organizers for a well-run and stimulating
conference, as well as the ATLAS and CMS Collaborations.
Special thanks go to Kamal Benslama and 
Georges Azuelos for their help with this manuscript.

\end{document}